\documentclass[twocolumn,preprintnumbers,amsmath,amssymb]{revtex4}

\usepackage{graphicx}
\usepackage{dcolumn}
\usepackage{bm}

\usepackage{epstopdf}

\newcounter{ichi}
\newcounter{ni}
\begin{document}


\title{Very-High-Energy Gamma-Ray Signal from Nuclear Photodisintegration \\
as a Probe of Extragalactic Sources of Ultrahigh-Energy Nuclei}

\author{Kohta Murase$^{1,2}$}
\author{John F. Beacom$^{2,3,4}$}
\affiliation{%
$^{1}$Yukawa Institute for Theoretical Physics, Kyoto University,
Kyoto, 606-8502, Japan\\
$^{2}$
CCAPP, The Ohio State University, 
Columbus, Ohio 43210, USA\\
$^{3}$
Department of Physics, The Ohio State University, 
Columbus, Ohio 43210, USA\\
$^{4}$
Department of Astronomy, The Ohio State University, 
Columbus, Ohio 43210, USA
}%

\date{July 1, 2010}
                        
\begin{abstract}
It is crucial to identify the ultrahigh-energy cosmic-ray (UHECR) sources and probe their unknown properties.
Recent results from the Pierre Auger Observatory favor a heavy nuclear composition for the UHECRs. 
Under the requirement that heavy nuclei survive in these sources, using gamma-ray bursts as an example, we predict a diagnostic gamma-ray signal, unique to nuclei --- the emission of de-excitation gamma rays following photodisintegration. 
These gamma rays, boosted from MeV to TeV-PeV energies, may be detectable by gamma-ray telescopes such as VERITAS, HESS, and MAGIC, and especially the next-generation CTA and AGIS.
They are a promising messenger to identify and study individual UHE nuclei accelerators.
\end{abstract}


\maketitle


\section{Introduction}
The origin of ultrahigh-energy cosmic rays (UHECRs) --- cosmic rays with energies above ${10}^{18.5}$~eV --- is one of the biggest mysteries in astroparticle physics, despite decades of efforts in theory and experiment~\cite{BS00, NW00}.
The Pierre Auger Observatory (PAO) and the High-Resolution Fly's Eye (HiRes) are providing newly precise clues, though these do not give a consistent picture.

The UHECR spectrum shows a cutoff around 60~EeV~\cite{PAO09s, HiR08}, consistent with the attenuation of ultrahigh-energy (UHE) protons due to photomeson interactions with the cosmic microwave background (CMB), but also consistent with the attenuation of UHE nuclei due to photodisintegration interactions with the cosmic infrared background (CIB)~\cite{All+05}. 
In either case, the spectrum at the highest energies must be replenished by sources within $\sim 100$ Mpc. 
The PAO data show a correlation between UHECR directions and nearby large-scale structure~\cite{PAO07}, which favors protons over nuclei, due to the smaller magnetic deflections; the HiRes data do not~\cite{HiRes10}.

Direct measurements of UHECR composition, through the average depth of shower maximum, $X_{\rm max}$, as well as the r.m.s. fluctuations around it, are now interestingly precise, and thus also probe the identities and properties of the sources.
This requires comparison to uncertain models of hadronic interactions, though the uncertainties are much less for the fluctuations. In both quantities, the PAO data favor a heavier composition above ${10}^{18.5}$~eV~\cite{PAO09c}; the HiRes data on $X_{\rm max}$ favor a proton composition~\cite{HiR09}.

To reconcile these inconsistent clues, not only UHECR observations but also other probes would also be needed, e.g., gamma rays and neutrinos that point back to their sources. 
This is especially true if UHECRs are nuclei. 
Their deflections in cosmic or Milky Way magnetic fields are larger than than those of protons, so that identification of sources seems more difficult (e.g., Refs.~\cite{TS09}).
The unprecedented precision of the PAO composition data, especially for the fluctuations around $X_{\rm max}$, motivates us to further consider a heavy composition. 
Though some heavy composition models have been suggested (e.g., Refs.~\cite{All+05}), most papers about gamma-ray and neutrino signals have focused on proton sources. 

Very-high-energy (VHE) gamma rays and neutrinos can be produced by sources of UHECR nuclei as well as protons.  
However, nuclei are special because they can produce nuclear de-excitation gamma rays following photodistintegration interactions. We show that this provides a characteristic energy spectrum that could differentiate extragalactic accelerators of nuclei and protons.


\section{VHE gamma-ray and neutrino production}
The most widely-discussed UHECR sources are gamma-ray bursts (GRBs)~\cite{Wax95,MINN08} and active galactic nuclei (AGN)~\cite{RB93, FG09}.
The predicted VHE spectra of gamma rays and neutrinos can be calculated with some assumptions, e.g., the yields from photomeson interactions depend on the cosmic-ray spectrum and target photon spectrum, and can strongly vary between source models 
(see, e.g., Refs.~\cite{MINN08, WB97, Mur07, AIM09, Mur09} for GRBs and Refs.~\cite{AD01, AHST08, Der+09} for AGN). 
GRBs and AGN can also be considered as sources of UHE nuclei~\cite{RB93,MINN08,WRM08,PMM09}.
In this work, as a demonstrative example, we mainly consider GRBs~\cite{Mes06}, and especially low-luminousity (LL) bursts such as GRB 060218, which might be more numerous than classical GRBs~\cite{Lia+07}.  
Applications to other sources such as AGN would be possible for given source models.

We assume a cosmic-ray energy spectrum of $\frac{d N_{\rm CR}^{\rm iso}}{d E_{A}} \propto E_{A}^{-p} {\rm e}^{-E_{A}/E_{A}^{\rm max}}$, where 
$E_A^{\rm max}$ is the maximum energy.  
We adopt $p = 2.3$, consistent with observations \cite{PAO09s, All+05}.  
The normalization is determined from the UHECR energy input rate, estimated as $E^2 \frac{d \dot{N}_{\rm CR}}{d E} \sim {10}^{44}~{\rm erg}~{\rm Mpc}^{-3} {\rm yr}^{-1}$ at $E_0={10}^{19}$ eV (e.g., Refs.~\cite{BS00}). 
For persistent sources, this is the product of the UHECR luminosity per source and source density.  
For transient sources, this is the product of the UHECR energy per burst $\tilde{\mathcal E}_{\rm CR}^{\rm iso} \equiv E^2 \frac{d N_{\rm CR}^{\rm iso}}{d E}$ and the apparent rate $\rho$.  
GRBs with $\rho \sim 1~{\rm Gpc}^{-3} {\rm yr}^{-1}$ require $\tilde{\mathcal E}_{\rm HECR}^{\rm iso} \equiv \tilde{\mathcal E}_{\rm CR}^{\rm iso}|_{E_0} \sim {10}^{53}$~erg~\cite{Wax95}.
LL GRBs, hypernovae and AGN flares, which may have $\rho \sim {10}^{2-3}~{\rm Gpc}^{-3} {\rm yr}^{-1}$, require $\tilde{\mathcal E}_{\rm HECR}^{\rm iso} \sim {10}^{50-51}$~erg \cite{FG09, MINN08}. 

For the target photon spectrum in the source, we use a broken power law, $dn/d \varepsilon \propto \varepsilon^{-\alpha_{l,h}}$, as for the synchrotron emission mechanism.
Here $\varepsilon$ is the photon energy in the comoving frame, $\varepsilon_{\rm ob} (\approx \Gamma \varepsilon)$ is that in the observer frame, and $\Gamma$ is the bulk Lorentz factor.  
For GRB  prompt emission, $\alpha_l \sim 1$ for $\varepsilon < \varepsilon^b$ and $\alpha_h \sim 2$ for $\varepsilon^b < \varepsilon$ are typical values~\cite{Mes06}, and $\varepsilon^b$ is the break energy.

Both protons and nuclei can undergo photomeson interactions above the pion production threshold.
For nuclei, the cross section and inelasticity are $\sigma_{\rm mes} \sim A \sigma_{p \gamma}$ (neglecting shadowing) and $\kappa_{\rm mes} \sim \kappa_{p \gamma}/A$~\cite{MINN08, AHST08}.
Using the $\Delta$-resonance approximation, the photomeson production efficiency
$f_{\rm mes} \approx t_{\rm dyn}/t_{\rm mes}$ is (e.g., \cite{WB97,MINN08})
\begin{equation}
f_{\rm mes} \simeq 5.5 \times{10}^{-4}
\frac{L_{\gamma, 46.2}^b}{r_{15.8} \Gamma_1^2 \varepsilon_{\rm ob, 5~keV}^b}
\left \{ \begin{array}{rl} 
{(E_A/E_{A b}^{(\rm mes)})}^{\alpha_h-1}\\
{(E_A/E_{A b}^{(\rm mes)})}^{\alpha_l-1} 
\end{array} \right. \,,
\end{equation}
where $E_{A b}^{(\rm mes)} \simeq 1.8 \times {10}^{17}~{\rm eV}~{(A/56)} {(\varepsilon_{\rm ob, 5~\rm keV}^b)}^{-1} \Gamma_1^2$ is the resonance energy, $L_{\gamma}^b$ is the photon luminosity at $\varepsilon_{\rm ob}^b$, $r$ is the emission radius, $t_{\rm dyn} \approx r/\Gamma c$ is the dynamical time scale of the relativistic source, and $t_{\rm mes}$ is the photomeson energy loss time scale.
Here, multi-pion production, which is important at sufficiently high energies, is not considered (c.f. Eq.~(18) in Ref.~\cite{MINN08}).

The value of $f_{\rm mes}$ has large uncertainties coming from parameters such as $r$. 
As in previous works~\cite{WB97, Mur09, AD01,AHST08}, one can consider cases where UHE nuclei are efficiently disintegrated in the source, i.e., where $f_{\rm mes} \sim 0.01-0.1$. 
But in this work, since we consider the PAO case where UHECRs are largely heavy nuclei, we focus on the scenario where GRBs are sources of UHE nuclei. 
As shown in Refs.~\cite{MINN08,WRM08}, besides several open issues such as particle escape, UHE nuclei rather than UHE protons are produced and survive in both the prompt and afterglow phases, when $r$ is large enough. 
We consider such cases where UHE iron can completely survive photodisintegration ($\tau_{A \gamma} \lesssim 1$) \cite{MINN08, WRM08, PMM09}, where $f_{\rm mes} \lesssim 1.9 \times {10}^{-3} {(A/56)}^{-1.21}$, and note that our purpose here is not to propose GRBs as sources of UHE nuclei.

Nuclei interact with target photons via the photodisintegration interaction, $A + \gamma \rightarrow {A'}^{*} + X$~\cite{Ste69}. The main contribution comes from the giant dipole resonance (GDR), which decays by the statistical emission of a single nucleon, though at higher energies, quasi-deutron emission and fragmentation occur.
After photodisintegration, the daughter nucleus is typically left in an excited state, which can immediately emit gamma rays as ${A'}^{*} \rightarrow A' + \gamma$. 
The gamma-ray multiplicity is $\sim 1-3$ and the energy in the nuclear rest frame is $\sim 1-5$~MeV. We adopt $\bar{\varepsilon}_{\rm Fe}=2.5$~MeV and $n_{\rm Fe}=2$~\cite{MF89}, so that the energy fraction carried by gamma rays is $\kappa_{\rm deex} \sim {10}^{-4} (56/A)$. 

The de-excitation process is not often taken into account for gamma-ray production.  Examples of application to Milky Way sources include Refs.~\cite{KKMT94, Anc+07}. 
Here we consider this process in an extragalactic context, which is relevant because we require that UHE nuclei from nearby sources survive sufficiently to reproduce the heavy UHECR composition deduced from the PAO data. 

In the GDR approximation, the de-excitation efficiency $f_{\rm deex} \approx t_{\rm dyn}/t_{\rm deex} \approx \kappa_{\rm deex} t_{\rm dyn}/t_{A \gamma}$ becomes (e.g., \cite{WRM08,PMM09})
\begin{equation}
f_{\rm deex} \simeq 2.8 \times {10}^{-5} \frac{L_{\gamma, 46.2}^b (A/56)^{0.21}}{r_{15.8} \Gamma_1^2 \varepsilon_{\rm ob, 5~keV}^b} 
\left \{ \begin{array}{rl} 
{(E_A/E_{A b}^{(A \gamma)})}^{\alpha_h-1}\\
{(E_A/E_{A b}^{(A \gamma)})}^{\alpha_l-1}
\end{array} \right. \,,
\end{equation}
where $E_{A b}^{(A \gamma)} \simeq 9.6 \times {10}^{15}~{\rm eV}~{(A/56)}^{0.79}  {(\varepsilon_{\rm ob, 5~\rm keV}^b)}^{-1} \Gamma_1^2$ is the resonance energy and $t_{A \gamma}$ is the photodisintegration time. 
The observed energy of de-excitation gamma rays from iron will be $E_{\gamma}^{(\rm deex)} \approx \Gamma \gamma_A \bar{\varepsilon}_{\rm Fe} \simeq 1.5~{\rm TeV}~{(A/56)}^{-1} E_{A,16.5}$.
Note that we have $f_{\rm mes} (E_{Ab}^{(\rm mes)}) \simeq 20 {(A/56)}^{-0.21} f_{\rm deex} (E_{Ab}^{(A \gamma)})$. 
We can see that the de-excitation efficiency is typically low when the nucleus-survival condition is satisfied, and the photomeson process is more efficient. 
Nevertheless, we see that the de-exicitation process leads to the interesting signal for nearby sources, and its component may dominate over the other components around the TeV scale.  

Gamma rays may be attenuated in the source by $e^{\pm}$ pair creation.
However, when the conditions are such that UHE iron survives in the sources (i.e., $\tau_{\rm Fe \gamma} \lesssim 1$), roughly speaking, we can neglect gamma-ray attenuation. 
This is because $\sigma_{\rm Fe \gamma} \simeq 8 \times {10}^{-26}~{\rm cm}^2 \sim 0.1 \sigma_{T} \sim \sigma_{\gamma \gamma}$ so that $\tau_{\gamma \gamma} \sim \tau_{\rm Fe \gamma}$. (Of course, note that the detail depends on the target photon field~\cite{MINN08}).
Gamma-ray attenuation also happens en route, even for the close sources we consider, and this is discussed below.

Neutrinos are also produced by nuclei.
The typical energy of conventional photomeson neutrinos is $E_{\nu}^{(\rm mes)} \simeq 0.28~{\rm PeV}~{(A/56)}^{-1} E_{A,17.5}$.
Neutrinos also come from the $\beta$-decay of nucleons from photodisintegration. 
A nucleon carries $\kappa_{\rm GDR} \sim 1/A$ of the energy of the nucleus, and about half of nucleons are neutrons.
The neutrino energy in the neutron rest frame is $\sim 0.48$~MeV~\cite{Anc+07}, so that
$E_{\nu}^{(\rm dis)} \simeq 0.29~{\rm TeV}~{(A/56)}^{-1} E_{A,16.5}$. 

\begin{figure}[t]
\includegraphics[width=1.00\linewidth]{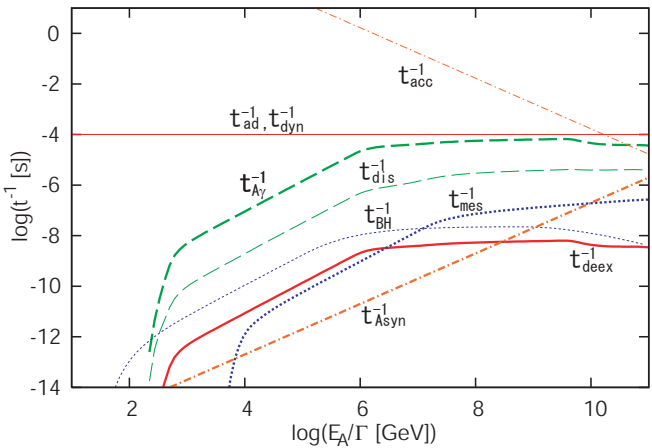}
\caption{\small{\label{Fig1}
The various cooling time scales and acceleration time for iron in the internal shock model for LL GRBs. 
Time scales are measured in the comoving frame of the outflow. 
The source parameters are $r={10}^{15.8}$ cm, $\Gamma=10$, $L_{\gamma}^b={10}^{46.2}~{\rm erg}~{\rm s}^{-1}$, $\alpha_l=1$~and~$\alpha_h=2.2$, and $\varepsilon_{\rm ob}^b=5$~keV~\cite{MINN08}. 
Note that the disintegration loss is usually dominated by emission of nucleons, i.e., $t_{\rm dis}^{-1}$ is larger than $t_{\rm deex}^{-1}$, $t_{\rm mes}^{-1}$ and $t_{\rm BH}^{-1}$.
Here, $t_{\rm dis}$ is the photodisintegration energy loss time, which is longer than the photodisintegration interaction time $t_{A \gamma}$, and $t_{\rm deex}$ is the de-excitation energy loss time due to gamma-ray emission. 
Note that $t_{\rm deex} \sim t_{\rm BH}$ at high energies, suggesting the potential importance of the de-excitation process. 
In fact, the de-exicitation process leads to production of higher-energy gamma rays, so that its component is more important at the VHE range in our case (see text).   
}}
\end{figure}

\begin{figure}[t]
\includegraphics[width=1.00\linewidth]{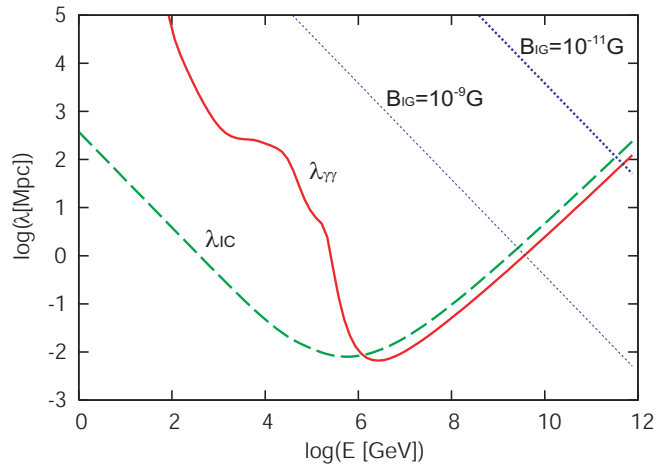}
\caption{\small{\label{Fig2}
The mean free path of high-energy photons for pair-creation and the energy loss length of electron-positron pairs for inverse-Compton in the Universe, respectively (solid and dashed lines). 
The dotted curves show the synchrotron cooling length for given IGMF strengths. 
}}
\end{figure}


\section{Predicted gamma-ray and neutrino fluences}
In the previous section, we analytically evaluated the nuclear de-excitation efficiency, and we can predict spectra of de-exicitation gamma rays from Eq.~(2). 
In this work, however, we perform much more elaborate calculations, using detailed cross sections of photodisintegration, photomeson, and Bethe-Heitler processes~\cite{Mur07,Ago+03}, where various cooling processes are taken into account.  
Following the method used in Refs.~\cite{Mur07, MINN08}, the photodisintegration time and other cooling time scales for nuclei and mesons (photodisintegration $t_{\rm dis}$, photomeson $t_{\rm mes}$, Bethe-Heitler $t_{\rm BH}$, synchrotron $t_{A\rm syn}$, inverse-Compton $t_{\rm IC}$, and adiabatic loss $t_{\rm ad}$) are numerically evaluated (see Fig.~1), and emissions by photomeson and cosmic-ray synchrotron processes are also calculated in detail. 
The magnetic field is determined from the equipartition between the magnetic energy density and the target photon energy density, which is often assumed in the classical optically synchrotron scenario for GRB prompt emission. 

The maximum energy $E_{A}^{\rm max}$ is estimated by comparing the acceleration time $t_{\rm acc} \approx 2 \pi \frac{E_A}{Z eBc \Gamma}$~\cite{WB97} with the dynamical and cooling times, and $E_{A}^{\rm max} \simeq (Z/26) {10}^{20.2}~{\rm eV}$ is obtained for our parameters described in the caption of Fig.~1. 
(Note that $t_{\rm acc}=\frac{E_A}{Z eBc \Gamma}$ is used in Ref.~\cite{MINN08}. 
We use a more conservative expression in this work.) 

We also use approximate formulae in Refs.~\cite{Anc+07} to evaluate primary spectra of gamma rays and neutrinos from photodisintegration.
In addition, the gamma-ray attenuation by pair creation with the CIB/CMB is considered. 
The mean free path of gamma rays for pair creation and energy loss length of electron-positron pairs for inverse-Compton are shown in Fig.~2.  

Motivated by the PAO results, we consider a heavy composition of 50\% iron and 50\% proton (to be conservative, not 100\% iron). The results for the de-excitation gamma rays would be unchanged as long as the composition is largely heavy nuclei of some species.
The condition $\tau_{\rm Fe \gamma} < 1$ and $\tau_{\gamma \gamma} < 1$ is satisfied for our adopted source parameters~\cite{MINN08}. 
Also, the corresponding baryon loading factors are ${\mathcal E}_{\rm CR}^{\rm iso}/{\mathcal E}_{\gamma}^{\rm iso} \sim {10}^{2-3}$.
Although such large values might be possible, smaller values which might be preferred can also be achieved for flatter or broken power-law cosmic-ray spectra~\cite{MINN08,AIM09}. 
The origin of the nuclei and the baryon loading depend on source physics and environments (e.g., Refs.~\cite{MINN08, WRM08} for GRBs).

The resulting gamma-ray fluxes are shown in Fig.~3.
At the highest energies, the primary photomeson gamma rays are dominant; however, these are severely attenuated by the CIB/CMB. 
Detections of the cosmic-ray synchrotron component (peaking at  $E_{\gamma}^{(A \rm syn)} \sim 33~{\rm MeV}~(Z/26) (A/56)^{-3} E_{A, 20}^2 B_{1.5} \Gamma_1^{-1}$) is also difficult, due to its insufficient flux. 
(The proton component typically seems more important in cosmic-ray synchrotron emission at higher energies, since $E_{\gamma}^{(A \rm syn)} \propto E_{A}^2 Z A^{-3}$~\cite{Ino10}).
In any case, not only nuclei but also protons, contribute to those meson and electromagnetic production processes. 
Hence, even if signals were detected, it would be difficult to prove that nuclei are accelerated in extragalactic sources.

The de-excitation process is invaluable as a probe of UHE nuclei accelerators. 
As in Eq.~(2), it has a characteristic spectrum in the VHE range, reflecting the target photon spectra, though the details depend on source models. 
This allows it to be distinguished from other processes, once the target photon spectra are known. 
The VHE range is doubly favorable: the attenuation due to the CIB/CMB is modest for nearby sources, and this is where the sensitivity of gamma-ray telescopes is best.

The fluence of de-excitation gamma rays shown in Fig.~3 can be reproduced with a simple estimate. 
Using Eq.~(2), we obtain $E_{\gamma}^2 \phi_{\gamma}^{(\rm deex)} \approx (f_{\rm deex} \tilde{\mathcal{E}}_{\rm CR, Fe}^{\rm iso})|_{E_{{\rm Fe} b}^{({\rm Fe} \gamma)}} / (4 \pi D^2) \sim 5 \times {10}^{-8}~{\rm erg}~{\rm cm}^{-2}$ at the break $E_{\gamma b}^{(\rm deex)} \simeq 0.5~{\rm TeV}~{(\varepsilon_{\rm ob, 5~keV}^b)}^{-1} \Gamma_1^2$. 
Although we have used typical parameters inferred from the observation of a LL GRB, one can explicitly see the parameter-dependence of the de-excitation gamma-ray flux from Eq.~(2).  
For GRBs at $D \lesssim 100$~Mpc, the de-excitation signals might be detected by present-generation gamma-ray telescopes such as VERITAS (Very Energetic Radiation Imaging Telescope Array System),  HESS (High Energy Stereoscopic System), and MAGIC (Major Atmospheric Gamma-ray Imaging Cherenkov Telescope). 
The prospects with next-generation telescopes such as CTA (Cherenkov Telescope Array) and AGIS (Advanced Gamma-ray Imaging System) are even more encouraging. 
HAWC (High Altitude Water Cherenkov observatory) (with TeV sensitivity of $\sim 5 \times {10}^{-7}~{\rm erg}~{\rm cm}^{-2}$ for $T = 5000$~s) would also be useful because of its very large field of view and duty cycle.

\begin{figure}[t]
\includegraphics[width=1.00\linewidth]{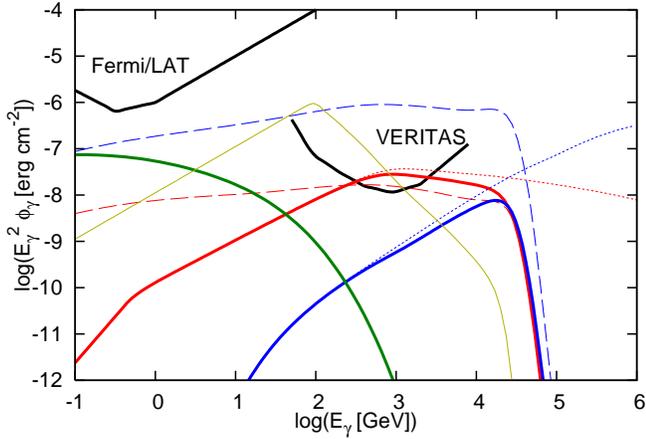}
\caption{\small{\label{Fig3}
Energy fluences of gamma rays from a LL GRB with
$\tilde{\mathcal E}_{\rm HECR}^{\rm iso}={10}^{50.5}$~erg and distance
$D=100$~Mpc.  
The source parameters are indicated in the caption of Fig.~1 with emission duration $T = 5000$~s.
The {\bf red curves} are the de-excitation component with attenuation
(thick solid line), without attenuation (thin dotted line), and with a possible
cascade component (thin dashed line). The {\bf blue curves} are the same
for the photomeson component, and the {\bf green curve} is the ion-synchrotron
component. For comparison, assuming that prompt emission is synchrotron, the SSC component is also shown (yellow curve). 
The sensitivities of Fermi/LAT~\cite{Fermi} and VERITAS 
(with a duty factor of 20~\%)~\cite{Car+08} are labeled.   
}}
\end{figure}

When gamma rays are attenuated by the CIB/CMB, the secondary electron-positron pairs created will upscatter CIB/CMB photons to gamma-ray energies, and the process repeats until the pair creation threshold is reached.
The detectability of this secondary cascade emission depends critically on the intergalactic magnetic field (IGMF). Especially, since the mean free path of TeV photons is hundreds of Mpc, pairs with $\sim$~TeV energies are likely to feel the IGMF in voids, which is very uncertain. 
(On the other hand, UHECRs should also feel the stronger IGMF in the structured region (clusters, filaments and sheets) and Milky Way magnetic field~\cite{Das+08}.)
If the IGMF in voids is strong enough ($B_{\rm IG} \gtrsim {10}^{-16}$~G), then the emission becomes nearly isotropic and its flux is diminished due to the magnetic time spread, so the cascaded gamma rays are not detectable~\cite{Pla95}. 

However, if the IGMF in voids is weak, there may be a detectable pair echo for transients or a pair halo for persistent sources~\cite{Pla95,ACV94}.
The echo duration is ${\Delta t}_{B} \sim {10}^{4}~{\rm s}~E_{e,12}^{-4} B_{\rm IG,-20}^2$, where $E_e$ is the energy of pairs~\cite{Pla95}.
The secondary cascaded gamma-ray spectra are shown in Fig.~3, which are calculated following Ref.~\cite{Mur09}. 
(Note that the calculations shown in Ref.~\cite{Mur09} were properly performed only for energies above the TeV scale, due to the focus on UHE gamma rays. 
Hence, the interaction and attenuation lengths of gamma rays and pairs in the Universe, shown in Fig.~1 of Ref.~\cite{Mur09}, are correct above the TeV scale.)
Thus, for a sufficiently weak IGMF, we expect that the cascaded photomeson gamma-ray flux, overwhelming the de-excitation gamma-ray flux, is detectable. 
However, it may be possible to discriminate between non-cascade and cascade components using temporal information~\cite{Pla95}.

\begin{figure}[t]
\includegraphics[width=1.00\linewidth]{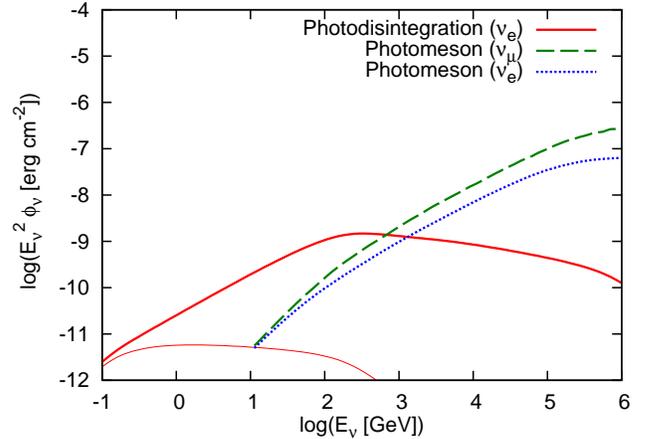}
\caption{\small{\label{Fig4}
Energy fluences of neutrinos from a LL GRB, as in Fig.~3.
For neutrinos from neutron decay, the time-integrated (thick solid line)
and early (within $T=5000$~s; thin solid line) cases are shown.  Neutrinos
from pion decay are also shown. Neutrino mixing is not taken into
account.
}}
\end{figure}

Depending on source models, there are possibilities for which the de-excitation component is not dominant in the VHE range. 
There would be synchrotron from pairs from decaying muons (originating from the photomeson process) and pairs from the Bethe-Heitler process. 
In GRB cases, photomeson production is typically more important than the Bethe-Heitler process at high energies ($f_{\rm mes} > f_{\rm BH} \sim {10}^{-4}$), and its energy fluence 
is $\sim (f_{\rm mes} \tilde{\mathcal{E}}_{\rm CR}^{\rm iso})|_{E_{A b}^{(\rm mes)}} {(E_\gamma/E_{\gamma b}^{(e \rm syn)})}^{\frac{1+\alpha-p}{2}}/8/(4 \pi D^2)$ with peak of $E_{\gamma b}^{(e {\rm syn})} \sim 1.1~{\rm GeV}~{(A/56)}^{-2} {(E_{A b,17}^{(\rm mes)})}^2 B_{1.5} \Gamma_1^{-1}$. 
The estimated fluence at TeV is $\sim 2 \times {10}^{-8}~{\rm erg}~{\rm cm}^{-2}$ for the pure iron case. 
Noticing $\kappa_{\rm BH} \sigma_{\rm BH} \approx 7.6 \times {10}^{-31}~{\rm cm}^2 (Z^2/A)$ at $\bar{\varepsilon} \sim 10 (2 m_e c^2)$~\cite{Ago+03}, the energy fluence of synchrotron emission by pairs from the Bethe-Heitler process is similarly estimated as $\sim (f_{\rm BH} \tilde{\mathcal{E}}_{\rm CR}^{\rm iso})|_{E_{A b}^{(\rm BH)}} {(E_\gamma/E_{\gamma b}^{(e \rm syn)})}^{\frac{1+\alpha-p}{2}}/(4 \pi D^2)$, where the effective optical depth for the Bethe-Heitler process is $f_{\rm BH} \simeq 0.76 \times {10}^{-4} {(Z/26)}^2 (56/A) L_{\gamma,46.2}^b r_{15.8}^{-1} \Gamma_1^{-2} {(\varepsilon_{\rm ob, 5~keV}^b)}^{-1} {(E_A/E_{Ab }^{(\rm BH)})}^{\alpha-1}$ and $E_{A b}^{(\rm BH)} \simeq 5.4 \times {10}^{15}~{\rm eV}~(A/56) {(\varepsilon_{\rm ob, 5~keV}^b)}^{-1} \Gamma_1^2$. 
The synchrotron peak is at $E_{\gamma b}^{(e {\rm syn})} \sim 1.3 \times {10}^{2}~{\rm eV}~{(A/56)}^{-2} {(E_{A b,15.5}^{(\rm BH)})}^2 B_{1.5} \Gamma_1^{-1}$, and the estimated fluence at TeV is $\sim 6.3 \times {10}^{-9}~{\rm erg}~{\rm cm}^{-2}$ for the pure iron case. 
The inverse-Compton emission by pairs, which is suppressed at high energies due to the Klein-Nishina (KN) effect, is also not dominant at the TeV scale. 
In our GRB cases, $\alpha_l \sim 1$ and $p \sim 2.3$ lead to detectable de-excitation signal, and we may expect that it dominates over the photomeson and Bethe-Heitler signals at the TeV scale. 
Generally speaking, however, the relative importance of the de-excitation signals depends on $B$, $\alpha$, $p$, and the composition.  

There may also be a purely leptonic component such as synchrotron self-Compton (SSC) emission. 
However, in our cases, the KN effect leads to significant suppression of SSC emission at very high energies. 
If keV-MeV emission is synchrotron from fast-cooling electrons, from $\varepsilon_{\rm ob}^b \sim$ keV and $U_B=U_{\gamma}$, the typical Lorentz factor of electrons is $\gamma_{e,i} \sim {10}^4$. The KN effect becomes important at $E_{\gamma \rm KN}^{(\rm SC)} \approx \Gamma \gamma_{e,i} m_e c^2 \sim 50$~GeV ($< E_{\gamma b}^{\rm (SC)} \approx 2 \gamma_{e,i}^2 \varepsilon_{\rm ob}^b$). 
Using the SSC-Y parameter in the Thomson limit, $Y_{\rm noKN} \approx (-1+\sqrt{1+4(U_e/U_B)})/2 \sim 0.6$, the SSC energy fluence at $E_{\gamma}=1$~TeV is estimated as~\cite{GG03}, $E_{\gamma}^2 \phi_{\gamma}^{(\rm SC)} \approx Y_{\rm noKN} (E_{\gamma}^2 \phi_{\gamma}^{(\rm syn)})|_{\varepsilon_{\rm ob}^b} {(E_{\gamma \rm KN}^{(\rm SC)}/E_{\gamma b}^{(\rm SC)})}^{2-\alpha_l}  {(E_{\gamma}/E_{\gamma \rm KN}^{(\rm SC)})}^{\alpha_l-p_e} \sim 2 \times {10}^{-8}~{\rm erg}~{\rm cm}^{-2}$ (see Fig.~3). 
Note that similar situations where the KN effect is significant are expected for GRB 080916C~\cite{GG03}.
But this is expected when the typical Lorentz factor of electrons is large enough, and resulting fluence depends on parameters such as $B$, $\Gamma$, $\tilde{\mathcal E}_{\rm HECR}^{\rm iso}$ and unknown GRB radiation mechanisms (which affect $\alpha$). 
For example, if $\gamma_{e,i} \sim {10}^{4}$, $\Gamma \sim 500$, and $\varepsilon_{\rm ob}^b \sim 250$~keV (which are expected for high-luminosity GRBs), the KN suppression occurs at $\sim 2.5$~TeV, while the de-excitation component peaks at $\sim 25$~TeV and may be visible around this VHE range.   
Therefore, the de-excitation and the other hadronic signals can dominate over the purely leptonic components at the VHE range, when the KN effect is significant. 
In our GRB cases, we expect that the de-excitation component may be visible above the TeV range (see Fig.~3), but the de-excitation and pionic gamma-ray components seem to be overwhelmed by electromagnetic (electron- and ion-synchrotron) components around the GeV range.   

Then, at least in our cases, the ``hard" de-excitation component is visible at $\gtrsim$~TeV over those from other processes in the sources or en route, though variations within the model uncertainties might change this. 
The purpose of this work is to demonstrate the potential importance of the de-excitation signals from UHECR sources, and to motivate VHE gamma-ray searches. 
Hence, here we avoid further studies, and expect that uncertainties will be reduced once sources are measured by future searches.

In Fig.~4, the neutrino fluences are shown.  
For IceCube detections, fluences of $\gtrsim {10}^{-4} E_{\nu,14}^{1/2}~{\rm erg}~{\rm cm}^{-2}$ at $\gtrsim 0.1$~PeV are required~\cite{WB97}. 
The time coincidence is important to reduce atmospheric neutrino backgrounds, but it would not work well for neutrinos from photodisintegration. 
Relativistic neutrons have lifetimes of $t_n \simeq \gamma_A~887~\rm s$ in the comoving frame, so that the fluence is suppressed at energies above $E_{\nu}^{\rm coin} \simeq 0.27~{\rm GeV}~{(A/56)} \Gamma_1^2 T_{3.7}$. 
Thus, when nuclei survive efficiently, we expect that their source
neutrino fluxes are too low to be detected.


\section{Summary and discussions}
We calculate the gamma-ray emission from UHE nuclei sources, inspired by the surprising PAO data suggesting that the UHECRs are largely heavy nuclei. 
We show that the de-excitation process of gamma-ray production, which is unique to nuclei, could be distinguished from other gamma-ray production processes and could be detected by VHE gamma-ray telescopes.
Therefore, it could identify extragalactic accelerators, with the detectable gamma rays probing cosmic-ray nuclei at the low end of the UHE range.
Importantly, our results show the most promising technique for identifying UHE nuclei sources, as the cosmic-ray nuclei will be strongly deflected~\cite{TS09}, the neutrino fluxes are low~\cite{MINN08,MB10}, and other hadronic gamma-ray signals such as pionic gamma rays cannot be cleanly identified.
If the nuclear signals are seen from UHECR sources, that would favor the PAO claim of a heavy nuclear composition. 
If they are not seen, then further measurements and studies will be needed to know the nature of UHECR sources. 
Obviously, all of the cosmic-ray, neutrino and gamma-ray experiments will be important (e.g., detecting UHE gamma rays~\cite{Mur09} and/or cosmogenic neutrinos~\cite{BZ69} would favor protons over nuclei).

Detections would be limited to nearby and/or energetic sources, because (1) the de-excitation efficiency is very low when iron nuclei survive, and (2) TeV-PeV photons from distant sources are attenuated by the CIB/CMB. 
If UHECR sources are transient, the event rate within 100~Mpc is $\sim 1.3~{(\tilde{\mathcal E}_{\rm HECR,50.5}^{\rm iso})}^{-1}~{\rm yr}^{-1}$~\cite{Mur09}.
These events can be found by low-energy all-sky monitors and followed up with VHE gamma-ray telescopes. 

Visibility of the de-excitation component may also be limited by other competing processes. 
As discussed in this work, the de-excitation process is important, but other photomeson, Bethe-Heitler, and purely leptonic processes such as SSC emission may be considered. 
Purely leptonic emission would be likely to be most important. 
If the baryon loading is not small and the KN suppression is significant, the hadronic signature may be observed at very high energies.  
Very high-energy pairs from the photomeson and Bethe-Heitler processes also lead to synchrotron and inverse-Compton emission. 
For $\alpha \sim 1$, the Bethe-Heitler energy loss rate of nuclei is roughly comparable to the de-exiction energy loss rate, but the typical energy of the de-excitation gamma rays is higher than that of gamma rays from pairs, which may make the de-excitaton signal observable (see above). 
Note that the photomeson energy loss rate is larger than that of the Bethe-Heitler energy loss rate at high energies in our cases (see Fig.~1), but the Bethe-Heitler process can be more important if photon indices are steeper (e.g., $\alpha \sim 2$).  
 
Despite these limitations, the most attractive feature of the nuclear de-excitation signal is that it is unique to nuclei and cannot be produced by protons. 
On the other hand, the photomeson and Bethe-Heitler processes are also induced by protons. 
The nuclear Bethe-Heitler process leads to more gamma rays than the proton Bethe-Heitler process for a $E_A^{-2}$ spectrum with given normalization. 
However, discrimination between protons and nuclei is not easy, unless we know each normalization of spectra of proton and nuclei. 
Generally speaking, all those three processes (de-excitation, photomeson, and Bethe-Heitler processes) can have important roles in UHE nuclei sources.  

Our calculations on the de-excitation signal are potentially important and formulae (see Eq.~(2)) are general. 
We have performed the detailed calculations for prompt emission from GRBs, but detectability is different for other models and associated parameters.
For example, one can apply our formulae to the external shock model for afterglows.
Although GRBs have been studied in this work as a detectable example, it would be possible to consider this process for other candidate UHECR sources, such as AGN and hypernovae.
As for AGN, though various acceleration and emission zones (e.g., blazar regions, hot spots and cocoon shocks) can be considered, the condition of nuclei surviving requires large emission radii, as discussed in Ref.~\cite{PMM09}. 
This means low photomeson, Bethe-Heitler, and de-excitation efficiencies, which limit our accessibility to UHE nuclei sources. 
In fact, the nucleus survival condition $\tau_{A \gamma}<1$~\cite{MB10} gives $f_{\rm mes} \lesssim 1.9 \times {10}^{-3} {(A/56)}^{-1.21}$, $f_{\rm BH} \lesssim 2.6 \times {10}^{-4} {(Z/26)}^2 {(A/56)}^{-2.21}$ and $f_{\rm deex} \lesssim 0.95 \times {10}^{-4} {(A/56)}^{-1}$.  
In the case of blazars, we typically expect strong SSC and/or external inverse-Compton emission, which may mask those weak hadronic signals.      
Radio galaxies with radio lobes and/or hot spots may also be viable candidates of nuclei~\cite{RB93}. 
Low efficiency of gamma-ray and neutrino production implies that detections of the nuclear signals may be possible only for nearby sources. 
In this sense, Cen A (at $\sim 3.8$~Mpc) may be of special interest. 
Some authors argued that all the UHECRs may come from Cen A as heavy nuclei~\cite{GTTT08}. 
Such the single source scenario can potentially be tested by constraining existence of the de-excitation and the other hadronic components.  


\acknowledgments
We thank C.~D. Dermer, S. Inoue, and T. De Young for useful information. 
K.M. acknowledges support by a Grant-in-Aid from JSPS and from CCAPP. 
J.F.B. is supported by NSF CAREER Grant PHY-0547102. 
The numerical calculations were carried out on Altix3700 BX2 at YITP in Kyoto University.

\clearpage





\end{document}